\documentclass[11pt, a4paper]{article}

\usepackage[paperwidth=210mm, paperheight=279mm,margin=2cm, marginparwidth=2cm]{geometry}

\usepackage{graphicx}

\usepackage{changepage}
\usepackage{amsmath,amssymb}

\usepackage{sidecap}

\usepackage{hyperref}
\providecommand{\href}[2]{\texttt{#2}}
\providecommand{\url}[1]{\texttt{#1}}

\newcommand{\eref}[1]{Eq.~\ref{#1}}
\newcommand{\fref}[1]{Fig.~\ref{#1}}
\newcommand{\Fref}[1]{\fref{#1}} 
\newcommand{\figref}[1]{\fref{#1}}
\newcommand{\sref}[1]{Section~\ref{#1}}

\newcommand{\tabref}[1]{Table~\ref{#1}}
\newcommand{\Tabref}[1]{\tabref{#1}}  

\providecommand{\citet}[1]{\cite{#1}}
\providecommand{\citep}[1]{\cite{#1}}

\newcommand{\tcomment}[1]{} 
\newcommand{\tnote}[1]{} 
\newcommand{\vnote}[1]{} 

\newcommand{\vvmatr}[1]{\mathbf{#1}}

\newcommand{\Gmatr}{\vvmatr{G}}
\newcommand{\Hmatr}{\vvmatr{H}}

\newcommand{\vvset}[1]{{\mathcal{#1}}}

\newcommand{\Ecal}{\vvset{E}}
\newcommand{\Gcal}{\vvset{G}}
\newcommand{\Hcal}{\vvset{H}}
\newcommand{\Kcal}{\vvset{K}}
\newcommand{\Ncal}{\vvset{N}}
\newcommand{\Pcal}{\vvset{P}}
\newcommand{\Rcal}{\vvset{R}}

\newcommand{\Dcal}{\vvset{D}}

\newcommand{\mathreg}{\mathrm{reg}}
\newcommand{\mathpop}{\mathrm{pop}}
\newcommand{\npop}{n^{(\mathrm{pop})}}
\newcommand{\Pcalpop}{\Pcal^{(\mathrm{pop})}}
\newcommand{\Pcalreg}{\Pcal^{(\mathrm{reg})}}
\newcommand{\Rbar}{\bar{R}}
\newcommand{\RbarM}{\Rbar^{(M)}}
\newcommand{\Rpop}{R^{(\mathpop)}}
\newcommand{\Rreg}{R^{(\mathreg)}}
\newcommand{\Rcalpop}{\Rcal^{(\mathpop)}}
\newcommand{\Rcalreg}{\Rcal^{(\mathreg)}}
\newcommand{\Rbarpop}{\Rbar^{(\mathpop)} }
\newcommand{\Rbarreg}{\Rbar^{(\mathreg)} }

\newcommand{\PR}{\mathrm{PR}} 

\newcommand{\emnote}[1]{\textit{#1}}

\newcommand{\nofive}{N${}^\mathrm{o}$5}
\newcommand{\perfume}[1]{``#1''}

\newcommand{\producer}[1]{#1}

\usepackage{hyperref}
\providecommand{\href}[2]{\texttt{#2}}
\providecommand{\url}[1]{\texttt{#1}}

\begin{document}
\vspace*{0.2in}

\begin{center}
{\Large\textbf{Social Success of Perfumes}} \\[6pt]
 {\large {V.\ Vasiliauskaite}\footnote{Corresponding Author.}},
 {\large \href{http://www.imperial.ac.uk/people/t.evans}{T.S.\ Evans}}
 \\[6pt]
\href{http://complexity.org.uk/}{Centre for Complexity Science}, and \href{http://www3.imperial.ac.uk/theoreticalphysics}{Theoretical Physics Group},
\\ Imperial College London, SW7 2AZ, U.K.
\end{center}
\section*{Abstract}
We study data on perfumes and their odour descriptors --- notes --- to understand how note compositions, called accords, influence successful fragrance formulas. We obtain accords which tend to be present in perfumes that receive significantly more customer ratings. Our findings show that the most popular notes and the most over-represented accords are different to those that have the strongest effect to the perfume ratings. We also used network centrality to understand which notes have the highest potential to enhance note compositions. We find that large degree notes, such as \emnote{musk} and \emnote{vanilla} as well as generically-named notes, e.g.\ \emnote{floral notes}, are amongst the notes that enhance accords the most. This work presents a framework which would be a timely tool for perfumers to explore a multidimensional space of scent compositions.



\section*{Introduction}

Smell is a cultural and social phenomenon. People (alongside other animals) bond over smell and associate odours perceived with certain memories \cite{AEGWPSHL13,HEBS04}. In some cultures, smell is so important that there are more adjectives to describe smells than there are for sights or sounds \cite{CHS94,G07}. Smell is an often undervalued yet potent emotional stimulant. Patrick S\"{u}skind in his book ``Perfume: The Story of a Murderer'' captivates not only with an engrossing story line but also with a power of smell over a man. The empowerment is well described in the following quote: ``Odors have a power of persuasion stronger than that of words, appearances, emotions, or will. The persuasive power of an odor cannot be fended off, it enters into us like breath into our lungs, it fills us up, imbues us totally. There is no remedy for it.'' \cite{S86}

In this work, we are interested in an artistic branch of olfaction  --- perfumery. Perfumery is the act of combining different olfactory ingredients, naturally occurring oils and chemical molecules, into a harmonious aromatic whole --- a perfume. For as long as records of perfumery have been kept, the first dating back to Mesopotamian times \cite{HEBS04}, the work of composing perfumes has been a job for ``the Nose'' --- an expert with the knowledge of pairwise complementary scent ingredients, their volatilities, odour longevities and other aspects that play role in perfume making. This expertise is typically acquired over many years of training and trials of many different combinations of ingredients.
This study explores the potential of on-line data to inform the art of perfumery by providing insights about the combinations of ingredients that lead to the most successful fragrance formulas.

%
%

A perfume is an exact chemical formula, developed by the Nose using his/hers years of experience of trial and error of multitudes of ingredient combinations. Each perfume constitutes of a specific combination of essential oils, which results in a unique scent of the perfume. It is then diluted with alcohol to result in \textit{cologne}, \textit{eau de perfume} or \textit{eau de toilette}.

Perfumes are often described using \textit{notes}. Notes are descriptors of scents that can be sensed upon the application of a perfume. Compositions of several notes, in particular the popular compositions that occur in many different perfumes, are called \textit{accords} (from the French for a musical \textit{chord}).

To create a well-balanced aromatic mixture, a variety of different smells are combined, so notes in a perfume are often varied and diverse. It is thought that a well-balanced perfume should comprise of ingredients with a wider range of volatilities: it should include some ingredients which evaporate quickly as well as those which linger for longer. This idea leads to a  classification of notes into one of three types: \textit{base notes} (least volatile), \textit{heart notes} (average volatility) and \textit{top notes} (most volatile) \cite{C62}.

Information of the precise amounts of each ingredient in the formulation of a perfume is confidential, to prevent duplications of the formula. However, the list of ingredients, the list of notes, is often advertised  in order to describe the scent of a perfume. Thus a perfume which smells of rose, vanilla and musk, is described using such notes.
In this study we have analysed the notes which make up over ten thousand perfumes without knowing anything about their specific amounts in each perfume. We assume that a note is included in the perfume description as its presence enriches the composition and its smell is detectable.

%
%

Most of the research on fragrances concerns biological and chemical features of olfaction\cite{BAAI08} and economics of perfume industry\cite{SKNG09}. Studies of human response to smell, such as how odours affect performance of certain tasks or mood have been conducted as well \cite{SSTFWGR12,SPCBM96,D05}. Olfaction is also part of the sense of flavour, alongside taste. Many studies explored how loss of smell influences the ability to sense flavours, for example see \cite{SABD10} and references therein.

In our work, we study perfumes and their constituent notes as a complex network.
Data driven approaches to market research and consumer trend analysis, for perfumes in particular, are now common.  For instance, artificial neural networks are now widely used in business and marketing where in the context of perfumes they have been used to identify customer requirements and to recommend future purchases to customers \cite{HRK10}.
However, perfume-note data has not been studied as a complex network. There are similarities with the analysis or food recipe networks\cite{KTN15,TLA11,TE17}, networks of flavour compounds\cite{AABB11,JRB15,VVWM13} and drug prescriptions\cite{CPBCSC12} as well as analysis of social media, such as Twitter\cite{AMW14}, concerning recipes.

%
%


Our work shows that our data on perfumes provides useful insights into the factors that are influential, and those which are not, when creating a successful product in the fragrance industry. We use positive and prolific customer feedback as out measure of success. We analyse multiple factors that could affect the observed success of a perfume: its launch date, popularity of its brand, price and ingredients. We compare potential success factors to popularity of perfumes as seen in an online database of perfumes.

We will assume that a large number of votes for a perfume is a measure of its success. This is a common assumption of most rating systems since in most cases voters leave positive feedback rather than criticise a product (for example see \cite{RZ02,M16b} especially the references and values in Table 30.1 of the latter).
In reality, there may be great perfumes that will never be highlighted as very popular. They may cater very well a small clientele, but not appeal to others due to their price, specificity or other factors. To account for this effect, we would need a much richer dataset that would include information about individuals reviewing the fragrances.
So in our study we assume that the larger the number of votes for a perfume, the more successful that perfume is which will inevitably penalise some great perfumes that are not universally popular.


%
%


\section*{Materials and methods}

%
%

\subsection*{Data}\label{SData}

We have information on 1047 different notes present in 10,599 perfumes. Users can provide a rating for each perfume and for each perfume $p$ we have the number of such `votes', $V_p$, and the average rating $R_p$. In addition the same web site also provided information about first year of production of each perfume. We also found prices for 978 of these perfumes since not all our perfumes are in production at the moment. In this study we consider prices in British Pounds per 100ml.

Our dataset required some cleaning. Some notes carried very similar names and we deemed these to synonyms for the same note. These differences could be due to spelling mistakes, the use of different languages or conventions. For instance, \emnote{Vanilla} (English) or \emnote{Vanille} (French) refer to the same note. In such cases, we would identify the two notes as identical and replace, for instance, all \emnote{Vanille} occurrences with \emnote{Vanilla}.
Another complication is that there may be notes with similar names whose odour profiles are distinct. For instance, our dataset contains \emnote{Vanilla}, \emnote{Tahitian Vanilla} and \emnote{Mexican Vanilla}, and the origin of an ingredient may determine its odour profile. We chose not to alter names of such special notes.
After this tidying, we were left with 990 notes, see \cite{VE18} for further information.



\subsection*{Methods}

%
%


For each perfume we have the number of votes and the average rating given by customers to perfumes; both these measures provide information about the success of the perfume.
The average customer rating can, however, be unreliable if it is based on a small number of votes. So it is useful to incorporate both the number of votes and the rating scores into a single effective rating. To do this we use a simple formula though one motivated by Bayesian statistics. Suppose that a perfume $p$ has an average rating of $R_p$ based on $V_p$ votes (votes). It is not unreasonable to compare this to $\RbarM$, the mean of the average rating of perfumes which have $M$ or more ratings.
Here $M$ is a parameter to be chosen but it is large enough such that we feel the ratings of individual perfumes with at least $M$ ratings are not unduly effected by the view of a few eccentric customers.
We then use a \textit{weighted score} $W_p$ defined as follows:
\begin{equation}
   W_p = \frac{V_p \, R_p + M \, \RbarM}{V_p+M} \, .
   \label{ebayesianmean}
\end{equation}
This can be derived in a Bayesian context assuming normal distributions for ratings as discussed in the Supplementary Information.
In our work we use $M=92$. This was chosen such that the mean number of reviews for perfumes with at least $M$ ratings was one standard deviation bigger than the mean number of reviews for all perfumes.

%
%



\begin{figure}[htb!]
\centering
\includegraphics[width=1\linewidth]{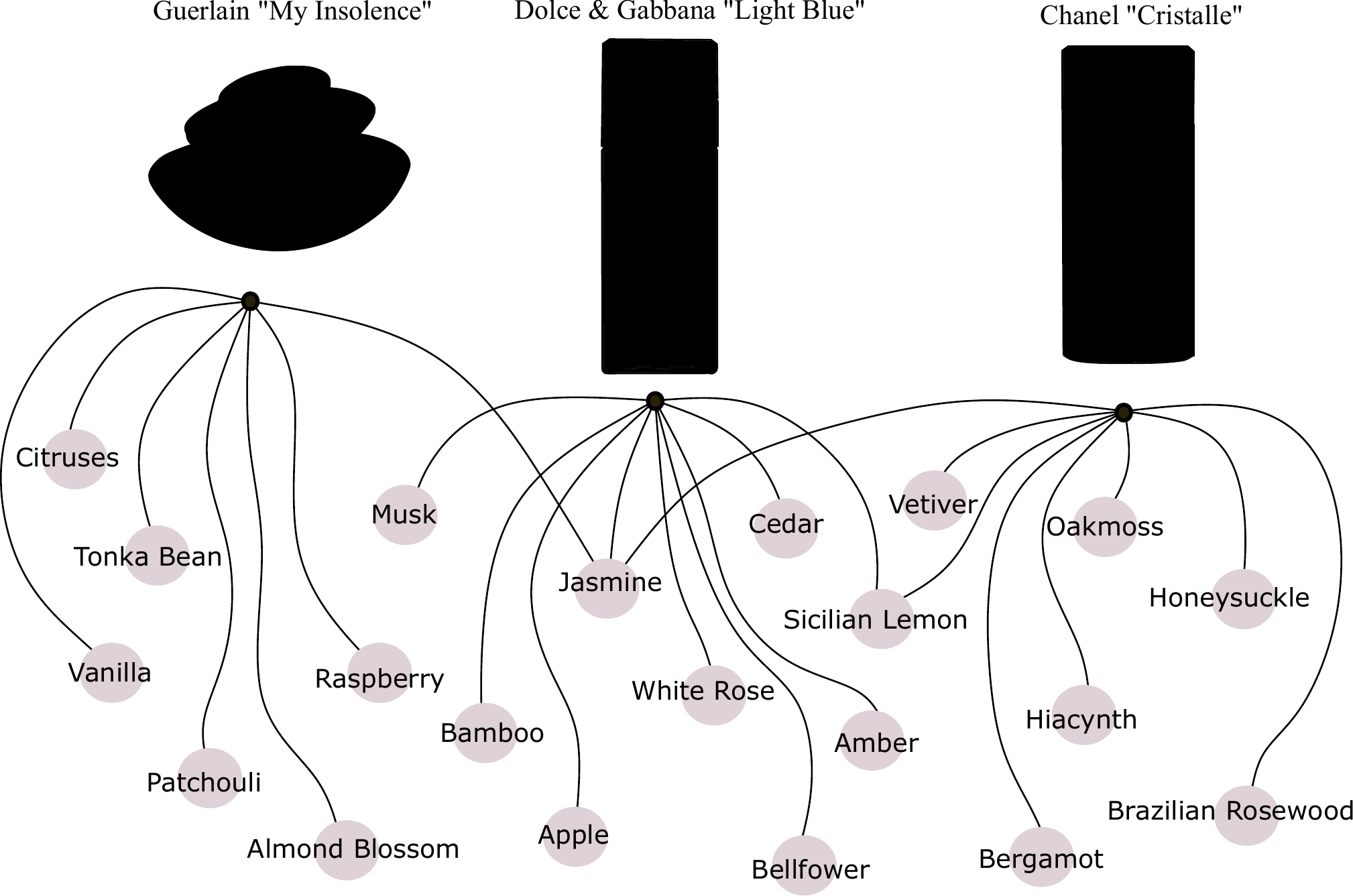}

\caption{An example of a perfume-note network $\Gcal$. An edge (black lines) is drawn between a perfume (a black dot with the perfume shown above it) and a note (large grey dots with names) only if that note features in the given perfume's composition.}
\label{fbipartite}
\end{figure}

To investigate how the success of a perfume is influenced by its note constituents, we use the network framework.
The most natural way to capture the relationships between perfumes and nodes in our data is to consider a \textit{perfume-note} network, $\Gcal$, in which we have two types of nodes: perfumes and notes. An edge is present between a note and a perfume only if that note is an ingredient of that perfume, making this a bipartite network.
An example of this network representation is given in \fref{fbipartite}.

%
%



\begin{figure}[htb!]
\centering
\includegraphics[width=0.8\linewidth]{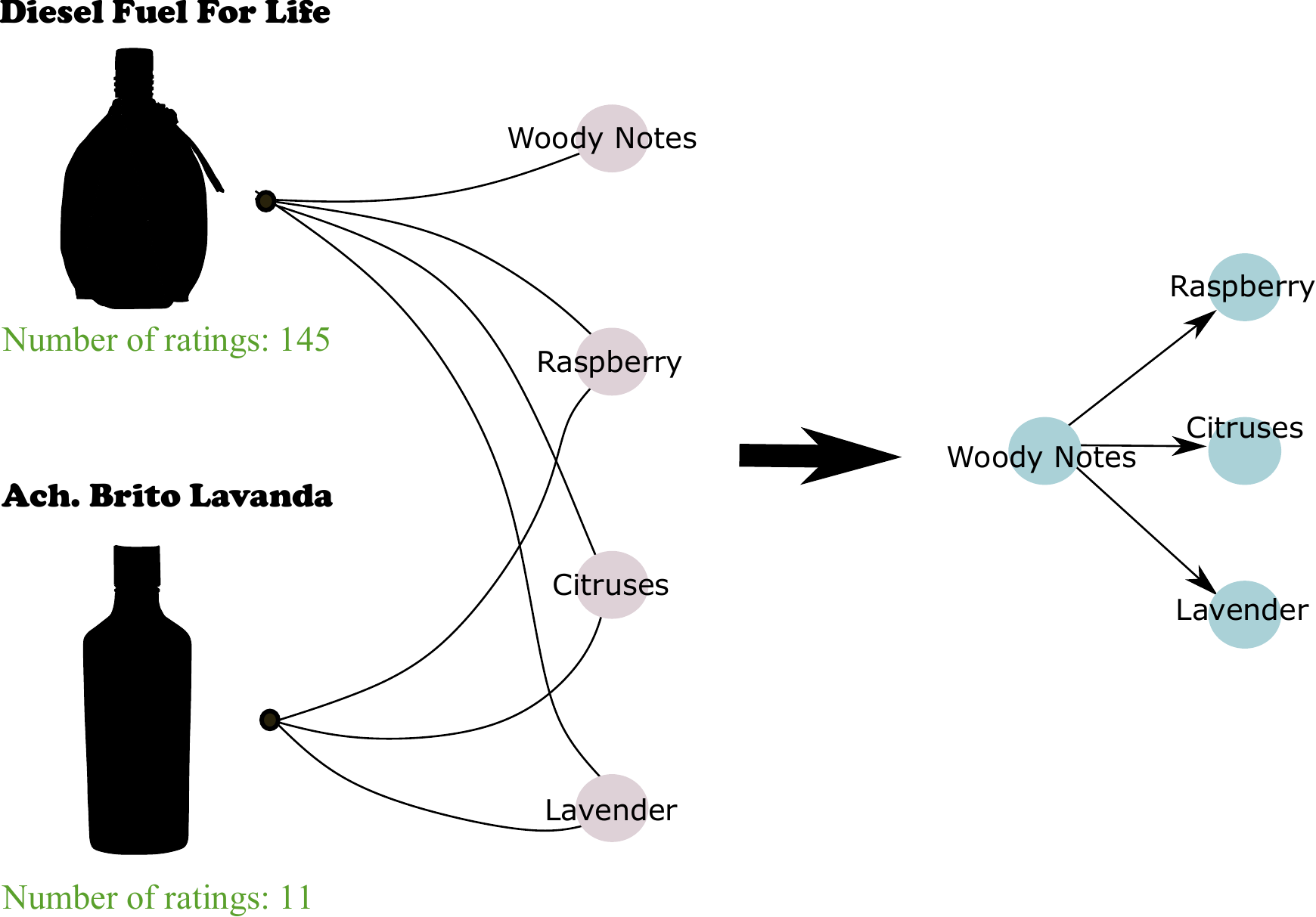}
\caption{An example of an enhancement event in perfumes. \textit{Woody Notes} are enhancing a composition of \textit{raspberry}, \textit{citruses} and \textit{lavender}. The last three notes feature in both ``Fuel for Life'' and ``Lavanda'' however, ``Fuel for Life'' has an additional \textit{woody notes} note and a higher number of reviews. Thus \textit{woody notes} must be enhancing the composition of \textit{raspberry}, \textit{citruses} and \textit{lavender}.}
\label{fenhancegraph}
\end{figure}

We also use a second network representation, a directed, weighted network which we will call an \textit{enhancement network} $\Hcal$. The nodes of this network are the notes, making this a type of one-mode projection of the bipartite network of perfumes and notes. However the definition of the weights and direction of the edges in our enhancement network is very different for other one-mode projections.
We start by setting the weight of all edges to be zero. We then look at pairs of perfumes where one has exactly one extra ingredient, which we call the \textit{difference note} $n_{\textrm{diff}}$, compared to the second perfume. If that is a positive enhancement, if the perfume with $n_{\textrm{diff}}$ has more reviews than the perfume with fewer ingredients, we assume that the addition of the extra ingredient to a set of notes is well thought out and that this one extra ingredient $n_{\textrm{diff}}$ has significantly enhanced the the overall composition. In that case we add one to the weight of a directed edge from note $n_{\textrm{diff}}$ to the nodes representing all the other notes in the two perfumes, as illustrated in \fref{fenhancegraph}.
By iterating through all possible pairs of perfumes, we form a weighted directed network in which a note has larger out-degree if it enhances many ingredients and larger in-degree if it has more potential to be enhanced.

We know of no other one-mode projection network which defines edges as in our enhancement network. Standard methods, such as those used in the context of other types of recipe, e.g.\ \cite{AABB11}, produce networks where edges are always reciprocated if not exactly symmetric, see \cite{ZK11} for an overview. By removing one set of nodes, any one-node projection of a bipartite network will always lose some information.  Likewise, by focussing on a relationships of pairs of notes, rather than a more complicated hypergraph representation, we may not encode all the relevant information available. However our aim with our enhancement network is to produce a representation of our data on perfumes which highlights key features while hiding aspects which are of little relevance. In particular, our use of metadata, here in the form of the votes, is designed to bring out important aspects of the data.
A more detailed definition and a discussion on possible variations of our enhancement network is given in the Supplementary Information.

\section*{Results}

%
%

\subsection*{Non-Network Results}\label{ssnonetresults}

\begin{figure}[htb!]
\centering
\includegraphics[width=0.7\linewidth]{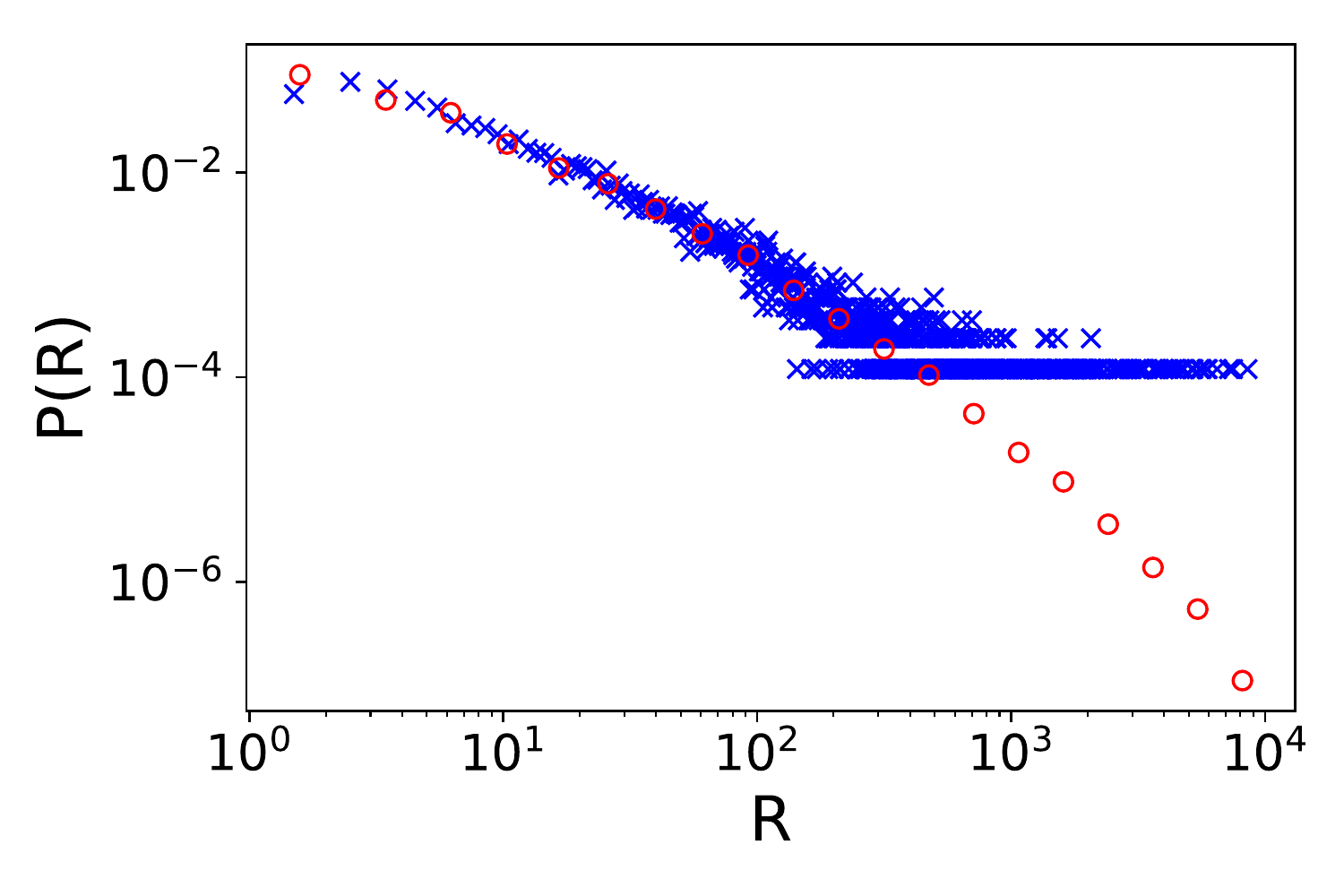}
\caption{Probability distribution of number of reviews R of perfumes. The real distribution of ratings (blue crosses) follows a fat-tailed distribution. The red circles show a logarithmically binned probability distribution which acts as a guide of eye to see that there are just a few perfumes which receive a large number of reviews.}
\label{fnoratingsdist}
\end{figure}

One measure of the impact or importance of a perfume is the number of reviews it has received, $V_p$. We find that the distribution of the number of reviews of perfumes is fat-tailed.  That is only a handful of perfumes receive a high number of reviews whereas the majority of perfumes receive little attention, see \fref{fnoratingsdist}.
Such fat-tailed distributions in the popularity of similar objects are common
as the degree distribution visualisations of the many data sets in the \href{http://konect.cc/plots/degree/}{Konect Project} \cite{konect} illustrate.
Using the number of reviews for each perfume, $V_p$, as a measure of their significance we find the top five to be, from largest to smallest $V_p$: \perfume{Light Blue} (\producer{D\&G}), \perfume{{J`adore}} (\producer{Dior}), \perfume{Euphoria} (\producer{Calvin Klein}), \perfume{\nofive} (\producer{Chanel}), and \perfume{Chloe} (\producer{Chloe}).

\begin{figure}[h!]
\centering
\includegraphics[width=1\linewidth]{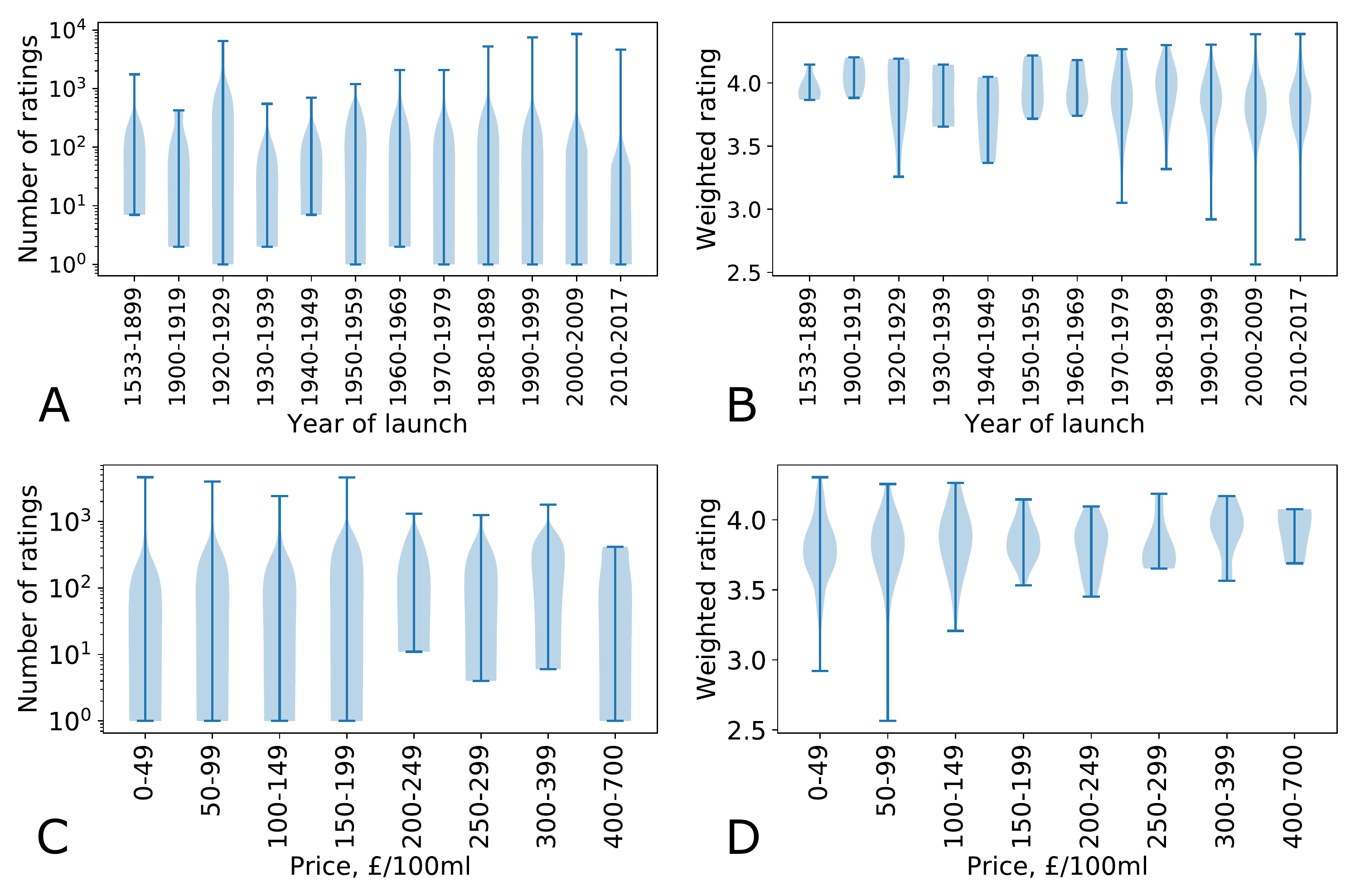}
\caption{Relation between popularity measures, the number of reviews and the normalised average score $W$, and either perfume launch date or price in {\pounds} per 100ml. Panels A, B show that the majority of older perfumes (launched before 1970s) have a relatively large normalised average score $W$, whereas there is a much larger variation in scores acquired for perfumes launched more recently.
Panels C and D show the relation between the two ratings and the price of perfumes. Perfumes that are of low price have a generally smaller number of reviews (the bell of a violin plot is concentrated close to 0) as opposed to more expensive perfumes, say those costing more than {\pounds}150/100ml or more. However, several perfumes that are cheap have a very large number of reviews. Panel D shows that the intervals of cheaper perfumes (price smaller that {\pounds}100/100ml) seem to be composed of a larger variety of perfumes: some with high score and some with low, whereas the more expensive perfumes have consistently high scores. Despite some differences in the spreads and distributions of $W$ and $V$ for perfumes in different age and price brackets, the figures do not reveal any strong correlation between the age or price of a perfume and its success.}
\label{fagepriceratings}
\end{figure}

On the other hand, the rating given by reviewers for any perfume is bounded (between 0 and 5) and the average rating value we have, $R_p$, is based on a sum of these values. So naturally, the distribution of these rating values is not fat tailed and they are typically clustered between 3.5 and 4.0 as is clear in \fref{fagepriceratings}. Clustering of ratings at high values is a common feature of ratings, for example see  \cite{HZP09}, since most ratings are positive \cite{RZ02,M16b}. 

We start by looking at the most successful perfumes to see if there are any common features. We began by studying the top-50 (roughly 5\%) of perfumes, based on number of reviews $V_p$ and by weighted score $W_p$.
After all, the price of a perfume covers many different costs, not just the ingredients. ``3\% of a perfume price is a smell'' \cite{T06}, the rest is packaging, advertising and margins. However, when we look at the top fifty lists, they contain perfumes which are very different.

One important factor in the success of a perfume can be its branding. As pointed out in \cite{T06}, there is a handful of companies, that constitute a majority of fragrance industry.
As expected, both lists of successful perfumes are dominated by well-known brands, such as \textit{Dior}, \textit{D\&G}, \textit{Chanel}, \textit{Nina Ricci}.
These brands may be more successful in the perfume industry because they have large revenues and monetary privilege enables such firms to create the best marketing campaigns.

The weighted rating $W_p$ highlights some cult perfumes, such as \perfume{\nofive} by \producer{Chanel}, \producer{Dior}'s \perfume{Poison}, and \perfume{Champs Elysees} by \producer{Guerlain}.

We also see classic vintage perfumes, some of which are no longer produced such as \perfume{Tabac Blond} by \producer{Caron} (released in 1922).
Celebrity perfumes also feature in the highly rated perfume lists, such as \perfume{White Diamonds} under the \producer{Elizabeth Taylor} brand (produced by \producer{Elizabeth Arden}). This is in agreement with a hypothesis that branding influences success of a perfume, as the name of a celebrity is a branding tool in and of itself.

Affordability can play a role as mid-range or even budget brands, such as \producer{L`Occitane} and \producer{Avon}, are also present in the lists of very popular perfumes. Their products being cheaper may well consist of lower quality ingredients.

What these lists of the top fifty most successful perfumes show is that none of the elements highlighted here, brand size, cult status, vintage classics, celebrity endorsement or price, seems to be the single determining factor in the success of these perfumes. This motivates us to look at the ingredients, using network methods, to see if these can throw light on what makes a successful perfume. Before that, we can look at the whole data set, not just the top fifty perfumes, to see if the age of a perfume or its price has an obvious effect on success.

We have both the age of a perfume (time since the launch date) and, in many cases, the price.
We have looked to see if there was any simple relationship between the age, price and the popularity of perfumes. To do this the data was binned, with wider bins for very old or very expensive perfumes where the data is sparse.

\begin{figure}[htb!]
\centering
\includegraphics[width=1\linewidth]{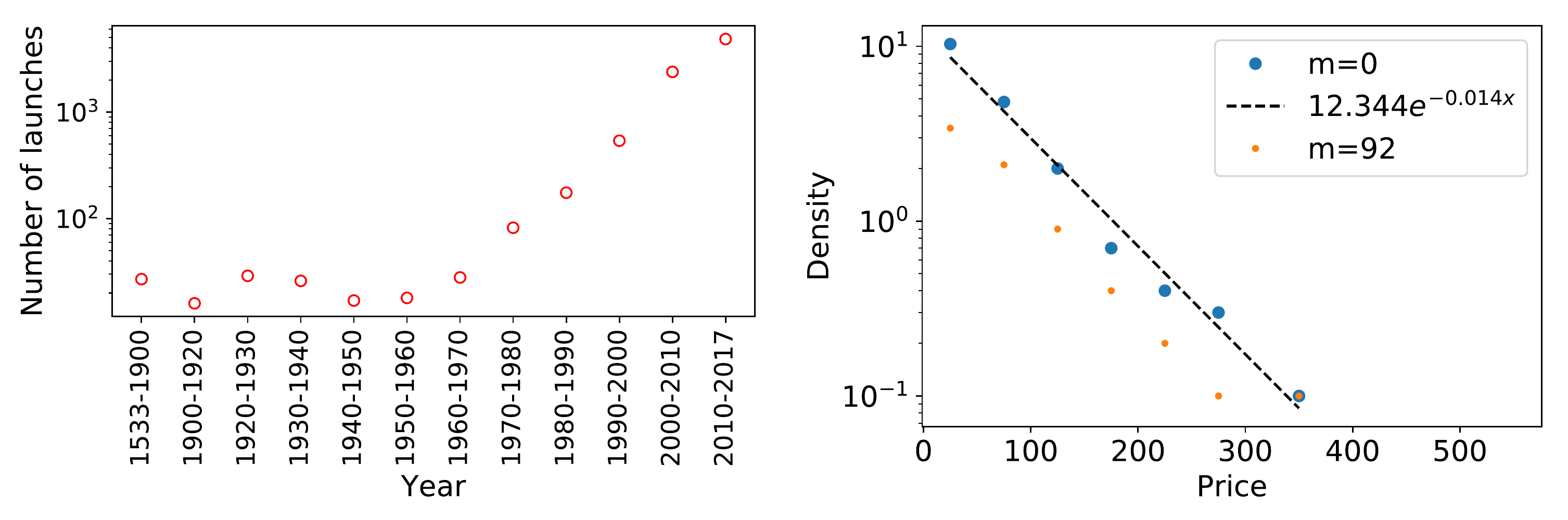}
\caption{On the left, the number of launches in each time period. Note the first two point cover more than a decade but all the others are decades. Note the roughly exponential rise from the 1950's. On the right, the number of perfumes launched in various price brackets.  The density of perfumes per bin is shown and these are plotted at the mid-point of the bin. Again the distribution falls off roughly exponentially. }
\label{fnolaunchesyear}
\end{figure}

Our database consists of 7635 perfumes with information about launch date. As seen in \fref{fnolaunchesyear}, the majority of perfumes in our dataset were launched relatively recently, around 95\% were launched in the last twenty years. In fact over the last sixty years, the number of perfumes with at least one rating in our data falls off roughly exponentially with age, $\sim \exp(y/9.9)$ where $y$ is the number of years since the perfume was launched, roughly 10\% less each year we go back.

There is also a small peak in the number of perfumes in our data which were first created in the 1920's and 30's.  This is when the first perfumes using artificial molecules were introduced creating the opportunity for both new sensations and for cheaper scents. The first perfumes to exploit this had a unique opportunity to create a fragrance with a large following that would then be some protection against similar examples created later.  This may explain why it is noticeable that perfumes created in this era are still discussed and even available today. The classic example here is \perfume{N${}^o$5} by Chanel which was the first perfume to use the synthetic compound `floral aldehyde', developed in 1921 by the famous perfumer Ernest Beaux.

\Fref{fnolaunchesyear} also shows that the number of perfumes also falls away very sharply with price as we would expect. Very roughly the number per price unit fell as  $\sim \exp ( v/70)$ where $v$ is the price in units of {\pounds} per 100ml.

The interesting question is to see whether there is any relationship between the age or price of a perfume and its success.  Our findings are visualised in \fref{fagepriceratings} (further tables are given in the Supplementary Information).

Panels A and B of \fref{fagepriceratings} show that there is little relation between perfume age and popularity, captured by either the number of reviews $V_p$ or the weighted score $W_p$.

The weighted rating varies more for the recent perfumes, where the older ones (created in the first quarter of 20th century and earlier) have more stable relatively high scores of around 4. This means that both the number of reviews and the average score of those ratings ought to be high for the old perfumes.  Perhaps the old perfumes withstood the test of time and are more likely to be universally acclaimed as high-quality perfumes, while the newer ones are much more varied in quality.

Panels C and D of \fref{fagepriceratings} show the relation between the price of the perfumes and their acquired popularity scores. Evidently, high quality and natural odourants are expensive, putting a high price tag on the resulting products. However, there seems to be little relation between the price of perfumes and their weighted ratings or the number of reviews received.
One explanation is that most people automatically take `value for money' into account in their rating, that is they normalise their rating to take account of the fact that they expect  more from an expensive perfume.
Another issue may be that different groups of people are rating cheap and expensive perfumes.
Such hypotheses would require a richer dataset than we have here, one which provided information on each reviewer (e.g.\ socioeconomic background) and the individual perfume ratings they have made.

So none of the factors discussed so far appear to be the sole key to the success for a perfume. Turin \cite{T06} when discussing the price of a perfume suggests that ``\ldots in fine fragrance there is a threshold below which a good fragrance is impossible, and we are probably there right now. However, more dosh does not necessarily mean better perfumes: some of the great fragrances of the past were relatively cheap formulae, and it is still quite possible to mix expensive raw materials and get an expensive mess''. So it appears that the choice of  ingredients and the way they are combined is vital for the success of a perfume so we now turn to study the notes used in perfumes.

%
%

\subsection*{Network Results}\label{SNetworks}



%
%

The popularity of notes, represented by their degree in the perfume-note network $\Gcal$, is not uniform. Indeed, we observed that some notes occur in the majority of fragrances while most notes are only used a handful of times (see Supplementary Information for a distribution). So if a note is used frequently does it have a better odour profile that tends to be preferred by customers and in turn makes perfumes containing that note more successful? In the perfume-note network $\Gcal$ these popular notes have more edges and thus have a higher degree. To investigate the influence of one popular note, $n_{\textrm{popular}}$, we will compare the rankings of perfumes with $n_{\textrm{popular}}$  and without.


Let $V_p$ be the number of reviews received by perfume $p$. We will then split the set of perfumes into two: one set of of perfumes $\Pcalpop$ all contain the chosen high-degree note $\npop$, while the remaining regular perfumes  without the popular note of interest are the subset $\Pcalreg$. We can then split the rating values (number of reviews) into two corresponding collections: $\Rpop$ with the ratings $V_p$ of perfumes containing the note $\npop$, and $\Rreg$ containing the ratings of the remaining perfumes.
If the ratings in the set of perfumes containing the popular note are higher than those in the regular set then we can deduce that that note has a positive effect on the success of a perfume. We do this by comparing the mean of the ratings in each set, $\Rbarpop$ and $\Rbarreg$.
To evaluate the confidence with which we can say that the average of one set is larger that the average of another we use two methods.

First we use Cohen’s $d$ score which is the difference between the means of two populations, normalised by the pooled standard deviation $s$\cite{C87}. That is
\begin{equation}\label{edscore}
 d = \frac{ \Rbarpop- \Rbarreg }{s} \, ,
 \qquad
 s = \sqrt{
 \frac{ (|\Rcalpop| -1 ) \sigma^2_{\mathpop} +
        (|\Rcalreg| -1 ) \sigma^2_{\mathreg}   }
      {  | \Rcalpop | + | \Rcalreg | -2 }
      }
 \,.
\end{equation}
Here $\sigma_{\mathpop}$ ($\sigma_{\mathreg}$) is the standard deviation of the ratings in the set $\Rcalpop$ ($\Rcalreg$).

We also used a permutation test with 10,000 permutations to look for significant effects of a popular note\cite{P37,R18}.  We use this to associate our $d$-score with a $p$-value which is the fraction of the random permutations which gave a larger $d$-score than found with the data. So a $d$-score with small $p$-value indicates that the effect seen in the data is significant as it is different from what would be found in the random case.
We saw little difference in the result when using a larger number of permutations and thus concluded that 10,000 trials suffices.

We only considered notes that featured in at least 100 perfumes with ratings where we might expect to have enough information to produce statistically significant result. The results for the ten most popular notes are summarised in \tabref{tpopularnotes}.  For these very popular notes, the perfumes containing these notes have a larger customer interest, $d>0$, but the effect is ``small'', $d\ll 1$. The $p$-values obtained from the permutation tests validate the significance of these results for all but two notes: \emnote{Bergamot} and \emnote{Mandarin Orange}, for which $p$-value is relatively large (larger than 0.01 which is a common confidence threshold).

\newlength{\tseEffectSize}
\settowidth{\tseEffectSize}{Effect}
\newlength{\tseDescriptor}
\settowidth{\tseDescriptor}{Descriptor}
\newlength{\tseNumberOfPerfumes}
\settowidth{\tseNumberOfPerfumes}{Number of Perfumes}
\newlength{\tseNumberOf}
\settowidth{\tseNumberOf}{Number of}
\begin{table}[h]
\centering
\caption{The ten most popular notes, their types (heart --- H, base --- B or top note --- T), degrees and effect on number of perfume ratings.  We say that the note is of specific type if most of perfumes list it as a note of this type (some notes are ``mobile'' in this sense: a note listed as, e.g.\ a heart note in one perfume may be listed as a top note in another). Note that this classification does not create hierarchy in notes: for instance, it is not clear whether the base note is hierarchically superior to top note. The last three columns contain information about how influential the note is for the number of reviews perfumes receive. The size of this effect on perfume ratings is calculated using $d$ of \eqref{edscore} (we used the standard notation to describe the effect size). To evaluate the validity of the result, we used the $p$-value of the permutation test. As the $p$-values show, we can confidently state their effect sizes except for \emnote{Bergamot} and \emnote{Mandarin Orange}. The effect sizes for the most popular notes are ``small'' at most. In our dataset, ``medium'' was the largest effect size of individual notes that was encountered. None of the top-10 most popular notes have such a large effect size.}
\begin{tabular}{c||ccc| c | c | c l | l }
\hline
 Note &
 \multicolumn{3}{c|}{Type} &
 \parbox[t]{\tseNumberOf}{\centering Number of \\perfumes\\ with note \\ \& ratings} &
 \parbox[t]{\tseNumberOf}{\centering Number of \\perfumes\\ with note} &
 \parbox[t]{\tseEffectSize}{\centering Effect \\ size \\ $d$} &
 \parbox[t]{\tseDescriptor}{\centering Effect \\ size \\ Descriptor } &	
 \parbox[t]{\tseDescriptor}{\centering Descriptor \\ $p$-value}
  \\ 
\hline \hline
Musk	    & B &   &   & 3839&	4768&	0.12	& Small	    & 0.0\\
Jasmine	    &   & H &   & 2562&	3239&	0.16	& Small	    & 0.0\\
Bergamot    &   &   & T & 2491&	3098&	0.05    & Very small& 0.043\\
Sandalwood  & B &   &   & 2421&	3063&	0.10    & Small     & 0.0\\
Amber	    & B &   & 	& 2321&	2892&	0.18	& Small     & 0.0\\
Rose	    &   & H &   & 2019&	2510&	0.11	& Small     & 0.0001\\
Vanilla	    & B &   & 	& 1442&	2397&	0.08	& Small     & 0.0012\\
Cedar	    & B &   & 	& 1387&	1895&	0.11	& Small     & 0.0004\\
Patchouli   & B &   &   & 1445&	1834&	0.09    & Small     & 0.0023\\
Mandarin Orange
         	&   &   & T & 1444&	1795&	0.05    & Very small& 0.07\\
\hline
\end{tabular}
\label{tpopularnotes}
\end{table}

On the other hand, we did find 60 notes with $p\leq 0.01$ associated with their $d$-score. In \tabref{tleffectnotes} we show the notes with the largest effect sizes showing clearly that these are not the ones used the most frequently (the most popular). From this we see that only five notes have more than a `small' effect on perfume ratings:\emnote{Anise}, \emnote{Orris Root}, \emnote{Orchid}, \emnote{Bamboo} and \emnote{Carnation}.

\begin{table}[h]
\begin{adjustwidth}{0in}{0in}
\centering
\caption{{Notes with the highest effects on perfume ratings. The note types are: H --- heart, B --- Base, or T --- Top. We only considered notes that were present in at least 100 perfumes (around 1\% of perfumes) and had $p$-value of the resulting d-score of no more than 0.01. We give Cohen's $d$ score and the descriptor in each case, along with a $p$-value assessing the significance of the description, so $p<0.01$ suggests the description is reliable.   We see that only five notes of our 990 have at least a moderate impact on perfume ratings: \emnote{Anise}, \emnote{Orris Root}, \emnote{Orchid}, \emnote{Bamboo} and \emnote{Carnation}.}}
\begin{tabular}{c||ccc| c | c | c l | l }
\hline
 Note &
 \multicolumn{3}{c|}{Type} &
 \parbox[t]{\tseNumberOf}{\centering Number of \\perfumes\\ with note \\ \& ratings} &
 \parbox[t]{\tseNumberOf}{\centering Number of \\perfumes\\ with note} &
 \parbox[t]{\tseEffectSize}{\centering Effect \\ size \\ $d$} &
 \parbox[t]{\tseDescriptor}{\centering Effect \\ size \\ Descriptor } &	
 \parbox[t]{\tseDescriptor}{\centering Descriptor \\ $p$-value}
  \\
\hline \hline
Anise   &   &   & T &  93 & 122 & 0.49 & Medium & 0.0005 \\
Orris Root	    &   & H &   & 206 & 220 & 0.46 & Medium & 0 \\
Orchid 	    &   & H &   &  266 & 323 & 0.41 & Medium & 0 \\
Bamboo	    &   & H &   & 84 & 102 & 0.37 & Medium & 0.0059 \\
Carnation 	    &   & H &   & 265 & 296 & 0.34 & Medium & 0 \\
Tuberose	    &   & H &   &  350 & 424 & 0.28 & Small & 0.0001 \\
Cyclamen 	    &   & H &   &  212 & 262 & 0.27 & Small & 0.0011 \\
Coriander 	    &   & H &   &  306 & 364 & 0.26 & Small & 0.0004 \\
Oakmoss    & B &   &   & 763 & 919 & 0.24 & Small & 0 \\
Lily 	    &   & H &   &  322 & 417 & 0.24 & Small & 0.0001\\
\hline
\end{tabular}
\label{tleffectnotes}
\end{adjustwidth}
\end{table}

So far we have looked at the effect of a single note on a perfume. However, perfumes contain combinations of notes, accords, which are carefully chosen. To illustrate, the example in \fref{fbipartite} shows an accord of \emnote{Jasmine} and \emnote{Sicilian Lemon} occurred twice, as this combination of notes features in two perfumes. An accord of \emnote{Vetiver} and \emnote{Honeysuckle} occurred once in \producer{Chanel}'s \perfume{Cristalle}, whereas an accord of \emnote{Musk} and \emnote{Vanilla} was not observed. If these two perfumes are successful, it might indicate that the \emnote{Jasmine}/\emnote{Sicilian Lemon} accord is an important aspect of that success. Searching for accords is analogous to a search of network motifs\cite{MSIKCA02}  in the perfume-note graph.

We are interested in the frequency of different accords, so we ask which accords occur in our dataset significantly more or less often than we would expect. To do this, we compare against a simple random model.
We have an `urn' containing the notes, every note appearing as many times as it does in our data set from the data (equal to the note's degree $k_n$ in $\Gcal$).
For every perfume in our data set, we now create a random version, drawing with replacement from the urn the same number of notes as the perfume had in the data (so the degree in $\Gcal$ $k_p$ is the same). We impose on restriction that no perfume can have the same note twice. Note that for each realisation, in which every perfume has been recreated using random notes, the notes used do not appear exactly as often as they do in the real data, but the average frequency of each note will be identical to the data.

To evaluate the significance of the frequency of an accord in our data we use a $z$-score and associated $p$-value. Suppose an accord occurs $f_{\textrm{real}}$  number of times in the data. We then measure the mean $\langle f_{\textrm{ran}}\rangle$ and the variance $\sigma_{\textrm{ran}}^2$ of the frequency of the same accord in our ensemble of random perfume-note combinations. Then the $z$-score of an accord is defined as
\begin{equation}
 z = \frac{f_{\textrm{real}}-\langle f_{\textrm{ran}}\rangle}{\sqrt{\sigma_{\textrm{ran}}^2}}.
\end{equation}
The $p$-value for the $z$-score of one accord is defined as the probability than that accord has a higher $z$-score in one of our random perfume-note combinations.

We can also calculate a $d$-score for the ratings of an accord in the same way as we did for a single note.  Now we create a set of rating values of perfumes which contain our chosen accord, $\Rcalpop$, and the ratings for the remaining perfumes go into $\Rcalreg$. The $d$-score of the accord, the size of the effect of the accord on the number of reviews of a perfume, is then given by \eref{edscore} as before. To determine significance of this $d$-score we use 10,000 permutations as before to find a $p$-value associated with this $d$-score.

To illustrate this, consider the two popular notes \emnote{Vanilla} and \emnote{Oakmoss} with high degrees in $\Gcal$: 2397 and 919, respectively.
As expected, these two notes were observed together as an accord in 145 real perfumes, which appears to be a large number. However, our null model shows they would be expected to occur together in around $224\pm15$ perfumes, giving a $z$-score of $-5.3$ and a $p$ value of 1. It means that the accord was more frequent in all  of our 1,000 random perfumes-note combinations (random networks) than it is in real data, i.e.\ this is statistically significant. So 145 perfumes containing \emnote{Vanilla} and \emnote{Oakmoss} is actually a significantly small number. We then say that such accord is under-represented, even though the combination was observed in over one hundred perfumes. We searched for all possible accords and evaluated whether they are over- or under-represented as well as whether they have an effect on the number of perfume ratings.


We counted the frequencies of accords (how often they occurred in the dataset) of two and three notes and compared them to the corresponding frequency in our null model. It allowed us to find both the over- and under-represented accords.
We set the following criteria when looking for accords whose over- or under-representing in the data was significant: the observed accord must occur in at least 1\% of perfumes, either $z>D_+=2$ or $z<D_-=0$, and the $p$-value is less than 0.01.

Using our criteria, we found 424 significant accords of size 2 with $z\geq 2$ and 764 significant accords with $z\geq 2$ of size 3.
The results of our findings are summarised in \tabref{tsigaccords}.

\begin{table}[h!]
\caption{{Table of accords which are over- and under-represented in the data (large $|z|$ values) and which also effect the number of reviews received by the perfumes in which the accords are present (large $d$ score). These accords also satisfy the criteria to appear in at least 1\% of perfumes and the $p$-value associated with the $z$-score is less than 0.01.
The first five accords (in italics) are those which are the most over- and under-represented in the data (largest $|z|$ values).
The remaining rows have the significant accords $z>2$ with the largest effect size ($d$-score) on the number of reviews of perfumes, at least 0.6 for accords of size two or 0.8 for accords of size 3. Such a large effect size means that perfumes which include these accords have a significantly larger number of reviews than you would expect.}}
\centering
\resizebox{\textwidth}{!}{
\begin{tabular}{ c || c | c | c | c |l |c|c}
\hline
Accords & Note  & $z$-score &	$z$-score & Cohen's   & Description &	$d$ score &	Number of \\
        & types &           &	$p$-value & $d$ score &             &	$p$-value & perfumes \\
\hline
\hline	
\textit{Ylang-Ylang, Aldehydes}&H T&17.8&0&0.51&Large&0.0002&130\\
\textit{Oakmoss, Carnation}&B H &15.9&0&0.58&Large&0.0001&121\\
\textit{Vanilla, Oakmoss}&B B&--5.3&1.0&0.34&Medium&0.0024&145\\

\hline

\textit{Ylang-Ylang, Aldehydes, Jasmine} & H T H & 32.99 & 0 & 0.55 & Large & 0.0001 & 117 \\
\textit{Amber, Musk, Jasmine} & B B H & 27.12 & 0 & 0.32 & Medium & 0 & 840 \\
\textit{Lily-of-the-Valley, Musk, Jasmine} & H B H & 25.43 & 0 & 0.32 & Medium & 0 & 495 \\
\textit{Lily-of-the-Valley, Rose, Jasmine} & H H H & 24.78 & 0 & 0.36 & Medium & 0 & 356 \\
\textit{Cedar, Geranium, Lavender} & B H H & 24.14 & 0 & 0.32 & Medium & 0.008 & 136\\

\textit{Musk, Vetiver, Vanilla}&B B B &-1.0&0.82&0.63&Large&0.0&132\\
 \hline    \hline
Amber, Orris Root & B H & 5.36 & 0 & 0.79 & Large & 0 & 110 \\
Sandalwood, Orris Root & B H & 6.67& 0 & 0.76 & Large & 0 & 124 \\
Jasmine, Orris Root & H H & 8.0 & 0 & 0.75 & Large & 0 & 139 \\
Amber, Coriander & B H & 2.62 & 0.002 & 0.73 & Large & 0 & 136 \\
Orris Root, Bergamot & H T & 7 & 0 & 0.72 & Large & 0 & 126 \\
Lily-of-the-Valley, Tuberose & H H & 6.70 & 0 & 0.66 & Large & 0.0001 & 111 \\
Lemon, Neroli & T T & 3.85 & 0 & 0.65 & Large & 0 & 128 \\
Musk, Orris Root & B H & 5.86 & 0 & 0.65 & Large & 0 & 154 \\
Rose, Orris Root & H H & 7.92 & 0 & 0.64 & Large & 0.0001 & 116 \\
Oakmoss, Lemon & B T & 9.02 & 0 & 0.63 & Large & 0 & 251 \\
Vanilla, Orchid & B H & 5.83 & 0 & 0.63 & Large & 0 & 136 \\
Sandalwood, Coriander & B H & 2.28 & 0.01 & 0.627 & Large & 0 & 137 \\
Melon, Peach & T T & 9.47 & 0 & 0.62 & Large & 0.0002 & 116\\
\hline
Amber, Oakmoss, Lemon & B B T & 11.09 & 0 & 1.13 & Very large & 0 & 120 \\
Oakmoss, Lemon, Jasmine & B T H & 10.60 & 0 & 1.08 & Very large & 0 & 124 \\
Sandalwood, Oakmoss, Lemon & B B T & 10.33 & 0 & 0.99 & Large & 0 & 116 \\
Amber, Oakmoss, Jasmine & B B H & 10.46 & 0 & 0.93 & Large & 0 & 186 \\
Cedar, Violet, Jasmine & B H H & 4.46 & 0 & 0.93 & Large & 0 & 108 \\
Amber, Neroli, Bergamot & B T T & 6.88 & 0 & 0.92 & Large & 0 & 107 \\
Amber, Oakmoss, Rose & B B H & 5.42 & 0 & 0.90 & Large & 0 & 119 \\
Musk, Jasmine, Orris Root & B H H & 14.13& 0 & 0.89 & Large & 0 & 112 \\
Oakmoss, Lemon, Cedar & B T B & 15.19 & 0 & 0.85 & Large & 0 & 116 \\
Cedar, Oakmoss, Amber & B B B & 11.00 & 0 & 0.82 & Large & 0 & 139 \\
Sandalwood, Jasmine, Neroli & B H T & 5.94 & 0 & 0.82 & Large & 0 & 107 \\
Oakmoss, Jasmine, Cedar & B H B & 9.70& 0 & 0.80 & Large & 0 & 1419\\
\hline
\end{tabular}
} 
\label{tsigaccords}
\end{table}	

\begin{figure}[h!]
\centering
\includegraphics[width=1\linewidth]{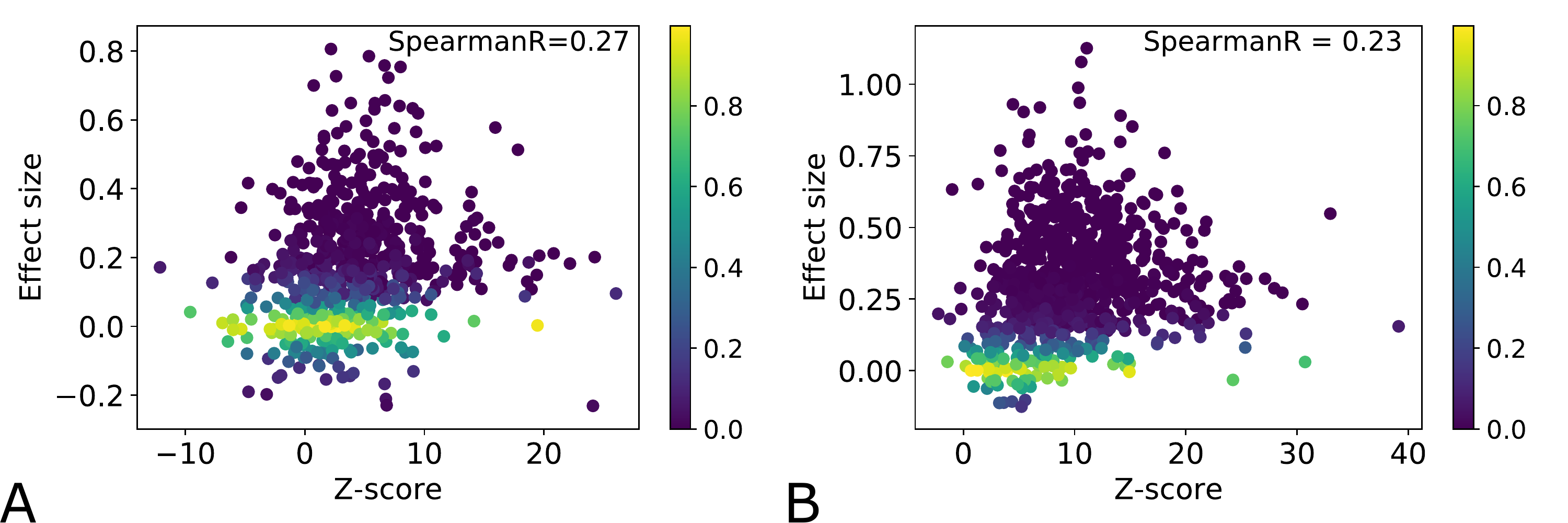}
\caption{Relation between $z$-score (over-representation) and effect size (influence on the number of reviews) for accords of size two (panel A) and size three (panel B). The two variables seem to be at best weakly related. The colour of a point indicates the $p$-value of the permutation test as shown in panel on the right of each plot.}
\label{fzscorevseffect}
\end{figure}

There is no clear relationship between $z$-score and $d$-score, as shown in \fref{fzscorevseffect}.  That suggests that simply using the most over-represented accords does not guarantee a successful perfume. There is, however, a significant number of  outliers, with either extreme $z$-values or with large $d$-scores.

The most over-represented accords ($z \gg 2$, see \tabref{tsigaccords}) seem to be composed of notes that are also very popular (see \tabref{tpopularnotes}), such as \emnote{Musk}, \emnote{Jasmine}, \emnote{Amber} and \emnote{Sandalwood}. There does not seem to be a common trend --- these most over-represented accords are not composed of polar opposite notes nor of very similar notes. Also we did not see any particular tendency to combine notes of similar nor different types (\textit{top}, \textit{heart} or \textit{base}).
The conclusion, therefore, is that these over-represented note combinations are indeed discovered by experimentation and multiple testing conducted by the `Nose'.

For instance, successful accords are not always made of notes of the same type. Two notes of different volatilities (different molecule sizes) may smell very similarly (share more of the odour compounds), and thus be more similar than some pairs of notes of the same type. Testing this idea further would require a richer dataset. At the same time, it can also be a good idea to combine notes with different smells. This happens in food as different cuisines can show a preference for similar tasting ingredients or they may combine ingredients that taste very different \cite{AABB11}. The musical analogy made for perfumes is again relevant as the notes  combined  can sound harmonious or dissonant and both can contribute to a successful piece.

However these accords which are most over-represented, those with large $z$-score such as shown at the top of \tabref{tsigaccords}, are not those with the largest effect on the number-of reviews (large $d$-score). This is also clear from \fref{fzscorevseffect}.

\Tabref{tsigaccords} shows the accords that have the most influence on the number of reviews. The most influential accords are: \emnote{Oakmoss}, \emnote{Lemon} and \emnote{Amber}; \emnote{Oakmoss}, \emnote{Jasmine} and \emnote{Lemon}; \emnote{Sandalwood}, \emnote{Lemon} and \emnote{Oakmoss};
\emnote{Amber}, \emnote{Oakmoss} and \emnote{Jasmine};
\emnote{Jasmine}, \emnote{Violet} and \emnote{Cedar}. Some examples of perfumes that consist of such accords are: \perfume{Eau Sauvage} by \producer{Christian Dior}, \perfume{\nofive} by \producer{Chanel}; \perfume{Acqua di Gio} by \producer{Giorgio Armani}; \perfume{White Diamonds} by \producer{Elizabeth Taylor}; \perfume{J'adore Dior} by \producer{Christian Dior}, \perfume{CK One} by \producer{Calvin Klein}. Thus our approach highlights perfumes that have high number of reviews $V_p$ as well as a weighted score $W_p$ by exploring the accord compositions that have strong effect size for the success of the perfumes.


We also looked at under-represented accords, finding 39 significant accords of size equal to two and one significant accord of size equal to three that have $z$-scores smaller than or equal to minus one and p-value larger or equal to $0.99$, see \tabref{tpopularnotes}.
We were able to distinguish some interesting structure for such under-represented accords as we noted some are composed of notes similar in nature, such as \emnote{Woody Notes} and \emnote{Sandalwood}, \emnote{Bergamot} and \emnote{Citruses}, \emnote{Lavender} and \emnote{Jasmine}. For instance, \emnote{Sandalwood} is a wood thus the two notes are wood-related scents; \emnote{Bergamot} has a citrus smell, so is similar to citruses. One explanation could be that in a perfume we sometimes look for an interesting combination of a variety of diverse notes, rather than combine many similar notes so there is no point in using accords of very similar notes.
There are a few interesting examples, for instance, \emnote{Musk}, \emnote{Vetiver} and \emnote{Vanilla} seem to have a large effect size of $d=0.63$, yet is under-represented. Thus, perhaps some of the accords with negative $z$-scores indeed are potentially unexplored great combinations.

Our Enhancement network $\Hcal$ encodes which notes have the most positive influence on other notes.  As with all network analysis, we are assuming that a large number of shorter paths linking notes in our Enhancement network $\Hcal$ are indicating a strong relationship.  Se the Supplementary Information for further examples and discussion. Once that is accepted, we can use network centrality measures \cite{N10} which measure the importance of nodes in a network.
Note that $\Hcal$ is not a causal network and so it is not transitive: if \emnote{Musk} enhances \emnote{Vetiver} and \emnote{Vetiver} in turn enhances \emnote{Vanilla}, this does not mean that \emnote{Musk} enhances \emnote{Vanilla}. In this context centrality value of \emnote{Musk} is related to its potential to enhance any composition of notes from $\Hcal$.
This type of importance is well measured using out-degree centrality, closeness centrality defined in terms of outgoing paths and reversed PageRank (PageRank applied to the enhancement graph with edge directions reversed). Out-degree, the number of edges pointing away from the note, tell us how many different notes the note enhanced. Furthermore, the weighted out-degree gives information about how many of the enhancing events (a note enhanced another note) were observed. Out-closeness centrality of a note shows the global effect of a note as an enhancer of a composition. The larger the out-closeness score of a note, the more it is likely to enhance other notes in the enhancement network. Lastly, PageRank counts how many edges are pointing to the note and the quality of those edges. Since we are interested in the outward edges, for this work we are reversing the edge direction when applying PageRank. We give the  definitions, mathematical formulae, and interpretations of the centrality measures used in this work in our Supplementary Information.

The resulting enhancement graph network has 165 nodes with 530 edges, whose total weight is 1423 --- the number of enhancing events. The largest weakly connected component contains 163 nodes and 529 edges (weight is 1422). The largest centrality notes and their centrality scores are summarised in \tabref{tcentralitytop}.  We saw little difference in results for different PageRank parameter $\alpha$ (see Supplementary Information) values between 0.7 and 0.95 so we show results for the traditional value of 0.85.

\newlength{\tseNote}
\settowidth{\tseNote}{Lemon Verbena}
\newlength{\tseScore}
\settowidth{\tseScore}{$0.99999999$}
\newlength{\tseNoteScore}
\setlength{\tseNoteScore}{\tseNote}
\addtolength{\tseNoteScore}{\tseScore}
\begin{table}[h!]
\begin{adjustwidth}{0in}{0in}

\centering
\caption{{Notes with the highest centrality scores in the Enhancement network $\Hcal$. Detailed definitions of these centrality measures are given in the Supplementary Information. The largest connected component of $\Hcal$ was used to calculate centrality.}}
\label{tcentralitytop}
\begin{tabular}{ p{\tseNote} | p{\tseScore}  || p{\tseNote} | p{\tseScore}  || p{\tseNote} | p{\tseScore}  }
\hline
\multicolumn{2}{p{\tseNoteScore}||}{\centering Out-strength} &
\multicolumn{2}{p{\tseNoteScore}||}{\centering Out-closeness \\ {(weighted, WF improved)} } &
\multicolumn{2}{p{\tseNoteScore}}{\centering Reversed PageRank \\ $\alpha=0.85$}  \\
Note	& Score &	Note&	Score	&Note	&Score\\
\hline \hline
Woody Notes & 180 & Musk & 0.252 & Floral Notes & 0.039 \\
Musk & 106 & Green Notes & 0.238 & Woody Notes & 0.031 \\
Floral Notes & 89 & Amber & 0.233 & Calone & 0.027 \\
Vanilla & 70 & Lavender & 0.226 & Orris Root & 0.024 \\
Fruity Notes & 68 & Bergamot & 0.226 & Sand & 0.023 \\
Citruses & 65 & Citruses & 0.218 & Musk & 0.023 \\
Spicy Notes & 64 & Spicy Notes & 0.215 & Jasmine & 0.023 \\
Frangipani & 44 & Lotus & 0.214 & Salt & 0.02 \\
Jasmine & 37 & Lemon & 0.213 & Amber & 0.019 \\
Amber & 35 & Lemon Verbena & 0.209 & Fig & 0.019\\
\hline
\end{tabular}
\end{adjustwidth}
\end{table}

Notes, with the highest enhancement effect fall into two categories. First, the high degree notes (\emnote{musk}, \emnote{vanilla}, \emnote{jasmine}) generally tend to enhance the composition. This is quite expected, as perhaps due to their universality they are popular notes to use in perfumery. Secondly, the list is dominated by generic notes, such as \emnote{woody notes} or \emnote{green notes}. Perhaps these are the ingredients that are not publicly disclosed, some ``secret formulas'' that make perfumes more complex and give depth to compositions.

\section*{Conclusion}

In this work we studied on-line data about fragrances to understand what makes a successful perfume. We found that the launch date and price correlates little to the popularity of perfumes. However, we did see major fashion brands were highlighted amongst producers of the most successful perfumes in the dataset. We further studied the structure of perfume-note bipartite network $\Gcal $ to understand the most over-represented combinations of notes of size two and three. We discovered that notes that are generally popular (have high degrees in $\Gcal$) also feature in the most over-represented accords. The most over-represented accord of size two is composed of \emnote{Geranium} and \emnote{Lavender}; accord of size three is \emnote{Oakmoss}, \emnote{Geranium} and \emnote{Lavender}. We were unable to see any simple tendencies in the most used accords, for instance neither accords of the same type (based on volatility) nor of different types seem to be favoured, so the experts are finding harmonies in their accords that transcend the basic data we have on each note.

There are a few under-represented accords, which could just be poor combinations.  However, two of them,  \emnote{Jasmine}/\emnote{Mint} and \emnote{Musk}/\emnote{Vetiver}/\emnote{Vanilla} do have a large positive effect on perfume ratings. Our results suggests these accords should be more popular than they currently are and that they deserve more attention in the future.

To understand whether there is a correlation between popularity of accords and perfume success, we estimated the effect size on the number of reviews for accords of size two and three as well as individual notes. We found that the combinations with the strongest effect sizes are not the most over-represented. The largest effect sizes are that of accords of \emnote{Oakmoss} and \emnote{Lemon} with either \emnote{Amber} or \emnote{Jasmine}. So by using customer review and basic recipes for perfumes in terms of notes, our methods are able to retrieve the perfumes with high customer popularity scores, highlighting the accords which the experts have found to work well.

Lastly, we studied an enhancement network $\Hcal$ --- a directed weighted network of notes --- in which a directed edge points from one note to another if it seems to be enhancing a composition. We found that notes with the highest enhancing effects (based on their out-degree out-closeness centrality and reversed PageRank) are those generically named (e.g. \textit{floral notes}) as well as those of high degree (e.g.\ \emnote{musk}, \textit{vanilla}).

There are other well-known methods for studying collections of items in data, such as using $k-itemset$ analysis to produce association rules used to recommend additional items for customers to buy: notes are items, accords are itemsets, and perfumes are `customers'. In the simplest cases such analyses rely on the frequency of accord/itemsets but do not distinguish between different customers/perfumes. We found that in itself did not help in our analysis and in our approach we emphasise that our perfumes are very different, as denoted by the votes given to each one.

Our work provides insights into factors that play role in the success of perfumes. It also sets up a framework for a statistical analysis of fragrances based on simple properties and customer reviews. It could be a beneficial tool for systematic ingredient selection and act as an artificial \textit{Nose}.

\section*{Acknowledgments}
V.V. acknowledges support from EPSRC, grant number EP-R512540-1.


%
%
%


\appendix

\renewcommand{\thefigure}{A\arabic{figure}}

\setcounter{figure}{0}
\setcounter{table}{0}
\renewcommand{\thetable}{A\arabic{table}}

\section{Notation}

Notation, used throughout this work is summarised in Table \ref{tnotation}.
\begin{table}[htb!]
\begin{tabular}{r | l}
Notation & Description \\ \hline \hline
$\Ncal$ & The set of notes \\
$n,m$ & Notes, $n,m \in \Ncal$ \\ \hline
$\Pcal$ & The set of perfumes \\
$p,q$ & Perfumes, $p.q \in \Pcal$ \\ \hline
$\Kcal_p$ & The set of notes in perfume $p$ \\
$\Kcal_n$ & The set of perfumes containing note $n$ \\ \hline
M & \parbox{0.8\textwidth}{Perfumes with $M$ or more ratings are considered to have reliable ratings.\\ We chose $M=92$ for this work.
(was $m$)}\\
$V_p$ & The number of reviews (`votes') for perfume $p$  \\ 
$R_p$ & The average of the individual ratings for a perfume $p$ \\
$\RbarM$ & The mean average rating $R_p$ over all perfumes with at least $m$ ratings\\
$W_p$ & The weighted rating for perfume $p$ \\  
\hline
$\Gcal$, $\Gmatr$ & The perfume-note network and associated adjacency matrix \\
$\Hcal$, $\Hmatr$ & The enhancement network and associated adjacency matrix \\
\end{tabular}
\caption{Table of notation used in this paper}\label{tnotation}
\end{table}

There are many ways to quantify success based on customer ratings. First for each perfume $p$ we count the number of votes $V_p$. Each individual reviewer $i$ states if they `love', `like' or `dislike' a perfume, a score $R_{pi}$ which is not available to us. The web site converts this into an average rating $R_p$ between one and five for that perfume and this is the information we have.\tnote{Technically they rate each perfume as `love', `like' or `dislike'.  The web site converts these into scores of 5, 4 and 1 respectively.}

To perform permutation tests, we split the set of perfumes into two: one set of perfumes $\Pcalpop$ which contain a specified note $\npop$, while the remaining  perfumes  without the  note of interest are the subset $\Pcalreg$.
We can then create two collections of ratings $V_p$, $\Rpop$ ($\Rreg$) with the ratings $V_p$ of perfumes containing $\npop$ (without $\npop$).
\begin{eqnarray}
\Pcalpop = \{ p | ( p,\npop  )\in \Ecal\} \, ,
&&
\Pcalreg = \{ p | (p,n_{\textrm{popular}}  ) \not\in \Ecal\} \, ,
\\
\Rcalpop = \{ V_p | p \in \Pcalpop \} \, ,
&&
\Rcalreg = \{ V_p | p \in \Pcalreg \} \, ,
\end{eqnarray}
where $\Ecal$ is the set of edges in the perfume-note network $\Gcal$.

To avoid the problem that a perfume with one or two perfect ratings can dominate the lists of top perfumes, we use a standard formula to rescale the ratings producing our `weighted rating' $W_p$.  This uses the number of ratings $V_p$ given to a perfume and the mean rating for perfumes with a `reasonable' average ratings equal or larger than $M$, which is a parameter we have to fix.
So suppose $\Pcal^{(M)}$ is the set of perfumes with ratings equal to or greater than $M$
\begin{equation}
 \Pcal^{(M)} = \{ p | R_p \geq M  \}
\end{equation}
The mean average rating $\RbarM$ is then defined as
\begin{equation}
 \RbarM = \frac{1}{N^{(M)} }  \sum_{p \in \Pcal^{(M)}} V_p R_p
 \, ,
 \qquad
 N^{(M) }= \sum_{p \in \Pcal^{(M)}} V_p
\end{equation}
so $N^{(M) }$ is the total number of votes given to perfumes with at least $M$ votes each.

To avoid problems with perfumes of a few ratings, low $V_p$, having extreme average rating values $R_p$, we use then use a \textit{weighted score} $W_p$. This can be defined using a Bayesian interpretation.

Suppose that all ratings are normally distributed with the same standard deviation $\sigma$ but with different means, each mean corresponding to the rating for a perfume. It makes sense to choose our prior distribution to be a Normal distribution too but we are free to set the mean and standard deviation for this prior. A reasonable mean to choose is the average rating over all perfumes, $\RbarM$.  The standard derivation of this prior is going to be the expected error in this mean assuming we make $M$ measurements, i.e.\ $\sigma_\mathrm{prior} = \sigma/\sqrt{M}$. Here $M$ is not necessarily the total number of ratings for all perfumes and instead we treat $M$ as a parameter we can set. We have already stated that the ratings are assumed to be Normal distributed so the likelihood distribution is also a normal, with mean given by $R_p$, the current average rating for perfume $p$, and standard deviation for this average rating equal to $\sigma_p = \sigma /\sqrt{V_p}$ since we have $V_p$ votes for perfume $p$.  We want the  expected distribution for the current average rating $R_p$, the posterior distribution. Combining a normally distributed prior with a normally distributed likelihood distribution can be done exactly using Bayes theorem (for example see \citet{B17b}, Eq.\ 4.40) and we find that the posterior distribution is also a Normal distribution with mean given by the sum of means weighted by the inverses of the variances.  That is the expected mean of the posterior distribution is $W_p$ where
\begin{equation}
   W_p
   =
   \frac{(\sigma_p)^{-2} }{(\sigma_p)^{-2}+ (\sigma_\mathrm{prior})^{-2}} R_p
   +
   \frac{(\sigma_\mathrm{prior})^{-2}}{(\sigma_p)^{-2}+ (\sigma_\mathrm{prior})^{-2}} \RbarM
   =
   \frac{V_p \, R_p + M \, \RbarM}{V_p+M} \, .
   \label{abayesianmean}
\end{equation}
This weighted score (sometimes known loosely as a `Bayesian Rating') has often been attributed as being the basis for many online contexts such as the \href{https://www.imdb.com}{IMDb movie rating site}, but most modern sites do not disclose their current method.

In our work we use $M=92$ as this ensured that the mean number of votes for perfumes with at least $M$ ratings was one standard deviation bigger than the mean number of votes for all perfumes.

%
%
\section{Launch date}

We have a launch date for 7635 perfumes, about 72\% of the perfumes in our data. The dates have been binned, mostly in decades except for the two earliest bins where there are fewer data points.  The results are shown in \tabref{tage}. The majority of these perfumes in our dataset were launched relatively recently, with around 95\% launched in the last twenty years. Over the last sixty years, the number of perfumes with at least one rating in our data falls off roughly exponentially with age, $\sim \exp(y/9.9)$ where $y$ is the number of years since the perfume was launched, roughly 10\% less each year we go back\footnote{Alternatively the form $\sim \exp (7.1-y/8.5) + 1.0$ also gives a reasonable fit which illustrates the accuracy of these fits.}.

\begin{table}[htb!]
\centering
\begin{tabular}{c||c|c}
\parbox[t]{2.5cm}{\centering Launch date \\ interval} &
\parbox[t]{3.7cm}{\centering Number of perfumes \\ with ratings} &
\parbox[t]{3.7cm}{\centering Number of perfumes \\ with $\geq92$ ratings } \\
\hline \hline
1533-1899 & 27&8\\
1900-1919& 16&7\\
1920-1929&29&13\\
1930-1939&26&4\\
1940-1949&17&4\\
1950-1959&18&9\\
1960-1969&28&9\\
1970-1979&82&26\\
1980-1989&175&57\\
1990-1999&538&164\\
2000-2009&2386&609\\
2010-2017&4851&645\\ \hline
Total & 8193 & 1555
\end{tabular}
\caption{Popularity of perfumes by launch date. The data is sparse for very old perfumes so the time ranges are larger in that case. Otherwise, we combined perfumes into bins by decade.}
\label{tage}
\end{table}


%
%
\section{Price}

We found prices for 978 of our perfumes, about 9.2\% of the total and we quote results in British Pounds per 100ml.

The number of perfumes falls sharply as the increases, the number per price unit, $v$, falling roughly as $\sim \exp ( v/70)$. Bins are in units of \pounds 50/100ml except for the most expensive perfumes where larger bins were needed. Results are shown in \tabref{tprice}.
\begin{table}[htb]
\centering
\begin{tabular}{c||c|c }
Price interval & \multicolumn{2}{c}{Number of perfumes}
\\
\pounds / 100ml &	with ratings &	with $\geq 92$ ratings  \\
\hline \hline
0-49& 516&170\\
50-99&238&106\\
100-149&98&47\\
150-199&35&20\\
200-249&19&10\\
250-299&13&7\\
300-399&14&10\\
400-700&5&3\\ \hline
Total & 938 & 373
\end{tabular}
\caption{Price intervals used to bin perfumes. The data is sparse for very expensive perfumes so there the price range is increased. Otherwise, we combined perfumes into bins of width \pounds 50 or 100ml.}
\label{tprice}
\end{table}	

\clearpage
\section{Network Definitions and Properties}\label{app:netdef}

\subsection{Sets of vertices}

We use two networks in our work based on perfumes and notes as nodes. A perfume will be denoted with an index $p$ and the full set of perfumes is $\Pcal=\{p_1,p_2,...,p_{N_p}\}$, where $N_p=|\Pcal|$ is the total number of perfumes in the dataset. Similarly, notes are denoted with indices $n$ and the full set of notes is $\Ncal=\{n_1,n_2,...,n_{N_n }\}$ where $N_n=|\Ncal|$. The set of notes in perfume $p$ are denoted by $\Kcal_p$ while $\Kcal_n$ is the set of perfumes containing note $n$.

\subsection{Perfume-Note Network}

The \textit{perfume-note} network, $\Gcal$, is a bipartite graph. We have two types of nodes: perfumes $p \in \Pcal$ and notes $n \in \Ncal$. An edge is present between a note and a perfume only if that note is an ingredient of that perfume. The adjacency matrix, $G_{pn}$ of the perfume-note network is therefore
\begin{eqnarray}
G_{pn} = \left\{
		 \begin{array}{cc}
          1 & \mbox{if note $n$ is in perfume $p$} \\
          0 & \mbox{if note $n$ is not in perfume $p$} \\
         \end{array}
         \right.
         \, .
\end{eqnarray}

In this perfume-note network $\Gcal$, the neighbours of perfume $p$ is the set $\Kcal_p$ defined above, so the degree of the node associated with perfume $p$ is $k_p=|\Kcal_p|=\sum_n G_{pn}$. Likewise, in this network, the node the note $n$ has a set $\Kcal_n$ of perfumes as nearest neighbours and degree $k_n=|\Kcal_n|=\sum_p G_{pn}$.

\begin{figure}[h!]
\centering
\includegraphics[width=0.9\linewidth]{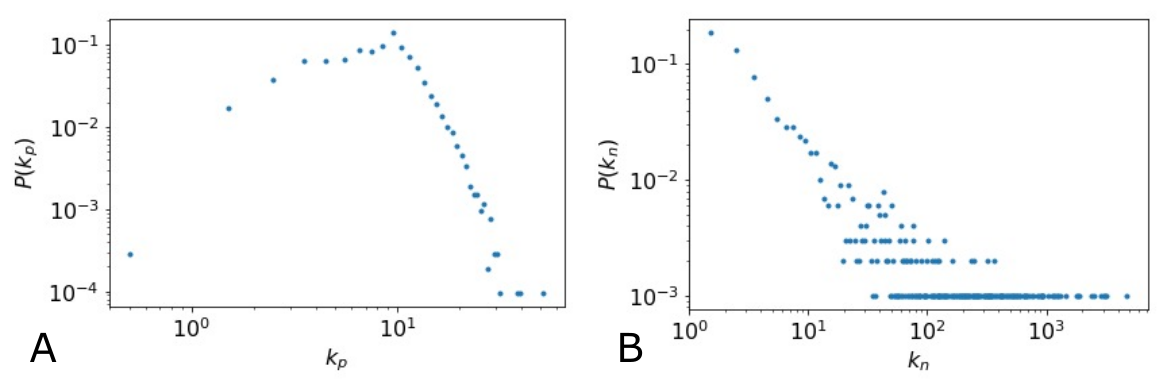}
\caption{Degree distribution of perfumes (panel A) and notes (panel B) in the perfume-note network $\Gcal$.}
\label{fddperfumenote}
\end{figure}

In our network, we have 10,599 perfume nodes and 990 note nodes with 89,388 edges. The network is given in \citet{VE18}. The degree distributions are shown in Fig. \ref{fddperfumenote}.

For the perfume nodes, we found that there are just seven perfumes that have ingredient lists of thirty or more perfumes while just 176 perfumes have twenty or more notes as their ingredients. By way of comparison, a normal distribution with the same number of perfumes having the ingredient lists of the same length as found in the data, we would expect just one perfume to have a list of length nineteen and none to have twenty or more notes in their ingredient lists.  So while the distribution of the degree of perfume nodes does have a noticeable tail for large $k_p$, it is not too extreme. This is to be expected as there is a limit to how many ingredients you can put into a single perfume and for them all to play a significant role.

The degree distribution for the note nodes is, on the other hand, clearly fat-tailed with an equivalent normal distribution giving nodes of degrees between about 55 and 125 only.  The most popular note is \emnote{Musk} used in 4768 perfumes, 44\% of perfumes, the tenth most popular \emnote{Mandarin Orange} is in 1795 perfumes (17\%). So it appears that like many other sets of similar objects (e.g.\ baby name popularity \citep{HB03}, dog breed popularity \citep{HBH04}), there is a `rich-get-richer' phenomena unlimited by any practical constraint leading to a fat tailed distribution in the note popularity.

%
%

\subsection{Enhancement Network}\label{SSEnhancement}

Our second network representation is a directed, weighted network which we call the \textit{enhancement network} $\Hcal$.

To understand why we create we first consider perfumes with exactly the same ingredient list
A couple of explanations for these come to mind.  First they could be almost identical in terms of their smell. However, the second possibility is that the concentrations of individual ingredients are not at all similar. These may govern the final smell of the perfume so this pair of perfumes may not smell the same in practice.
For instance, two perfumes can be composed of 3 ingredients: musk, rose and vanilla, however ratios are, 5:2:2 and 2:5:2. The first perfume ought to smell more ``musky'' and the second one more ``rosey''. However, the detailed compositions are invariably closely guarded secrets and as we do not have precise comparisons from users in our data, we can not distinguish these two cases . There is an obvious analogy here with food recipes.  The amount of chilli and the type of chilli used in a recipe can have a drastic effect on the user experience.  However, like us, most food recipe network studies do not include the quantities of ingredients in their analysis.

However, we can attempt to make deductions about notes through comparisons between similar perfumes in another way. We can look at pairs of perfumes where one perfume has exactly the same ingredients as the second perfume except for the addition of \textbf{one} extra note.
Our reasoning is that adding one extra note to a list of ingredients could ruin a perfume.
For instance, a small drop of violet could easily overpower the entire composition, despite the amount of it in the combination being small. However, we must assume that the expert `Nose' who created the perfume with an extra ingredient included for a good reason.  So if our rating for the perfume with the extra note is higher than the perfume without the extra note, we will assume that the extra ingredient \textit{enhances} the other notes.
We will assume that the addition of an extra ingredient to a set of notes is well thought of and significantly affecting the composition overall. Our enhancement network will encode these comparisons.

Of course we still do not know if in any two perfumes differing by one note the proportions of ingredients are similar.  So we will use a weighted network to capture how often we find a given enhancement.  In this way we will try to use the large amount of data to build up a statistically significant picture.

Formally, we define our enhancement network $\Hcal$ as follows. Each node is a note $n \in \Ncal$. To define the edges, consider two perfumes that are almost identical: one perfume $q$ has $k_q$ notes, the set of notes $\Kcal_q$, and the second perfume $p$ has the same $k_q$ notes plus one additional note which we call the \textit{difference note} $n_{\textrm{diff}}$. That is perfume $p$ contains the notes in the set $\Kcal_p = \Kcal_q \cup \{ n_{\textrm{diff}} \}$. Provided the number of votes of the first perfume is smaller than that of the latter, $R_q < R_p$, the note $n_{\textrm{diff}}$ must be enhancing the composition.  We therefore draw an edge from the difference note node $n_{\textrm{diff}}$ to the nodes representing the other notes in the two perfumes, the notes in common to both $\Kcal_q$. We then add one to the weight to each edge running from the node representing $n_{\textrm{diff}}$ to the other nodes, those in $\Kcal_q$.

The adjacency matrix may be written formally as follows
\begin{eqnarray}
 H_{mn}
 &=&
 \sum_{p,q \in \Pcal | p \neq q}
 G_{pn} G_{pm} G_{qm}
 \delta(\Ncal_p \cap \Ncal_q,\Ncal_q) \delta(  \Ncal_p \backslash \Ncal_q, n)
 \Theta(V_p,V_q)
 \, .
 \label{enhanceadjmat}
\end{eqnarray}
Here $\delta(A,B)$ is an indicator function which is 1 if $A=B$ and 0 otherwise. The notation is a cumbersome way of stating that perfume $q$ has notes $\Ncal_q$, containing note $m$, while perfume $p$ has the same notes plus one more, note $n$ ($n=n_{\textrm{diff}}$ the difference note), so that $\Ncal_p = n \cup \Ncal_q$. To enforce the requirement that the rating of perfume $p$ is higher than perfume $q$, $V_p>V_q$, we use the indicator function $\Theta(A,B)$ which is 1 if $A>B$ and 0 otherwise. Note the edge as described above is considered to be directed from note $n$ to note $m$.

Our enhancement network has 165 nodes (we ignored notes which were never involved in any enhancement), With 530 edges and a total weight of 1423.
The largest weakly connected component contains 163 nodes and 529 edges (total weight 1422).
The average shortest path length is 1.5 (accounting for the weights of edges).

Of course we could extend this, looking at perfumes which differ by different number of notes but in our opinion it is not so clear what is enhancing what in more complicated cases. Likewise our some examples differing by one note will have ratings ordered the other way round, $V_p>V_q$.  We might interpret the extra note as \textit{diminishing} the original recipe.  That could be captured in a \textit{diminishment network} $\Dcal$ with   adjacency matrix differing from \eqref{enhanceadjmat} only in the arguments of the indicator function $\Theta$
\begin{eqnarray}
 D_{mn}
 &=&
 \sum_{p,q \in \Pcal | p \neq q}
 G_{pn} G_{pm} G_{qm}
 \delta(\Ncal_p \cap \Ncal_q,\Ncal_q) \delta(  \Ncal_p \backslash \Ncal_q, n)
 \Theta(V_q,V_p) \, .
 \label{diminishadjmat}
\end{eqnarray}
It could be natural to consider these two networks together as a signed network with adjacency matrix $S_{mn} = H_{mn} - D_{mn}$.

\subsubsection{Enhancement Network Example}

In our work we apply centrality measures to our enhancement network, as discussed in \sref{ssenetcent} below. In order for such measures to have meaning, we requires that paths in the enhancement network have some meaning. That is two notes linked by an edge clearly have some sort of direct relationship, but most centrality measures also use longer paths to indicate a likely if weaker relationship.

\begin{SCfigure}[50][htb!]
  \centering
  \caption{A very simple Perfume-Note network $\Gcal_\mathrm{ex}$. Note that the vertical height of the perfume notes reflects the number of votes they get; the higher up the page a perfume node is, the more votes that node has.}
  \label{fpnex}
  \includegraphics[width=0.45\linewidth]{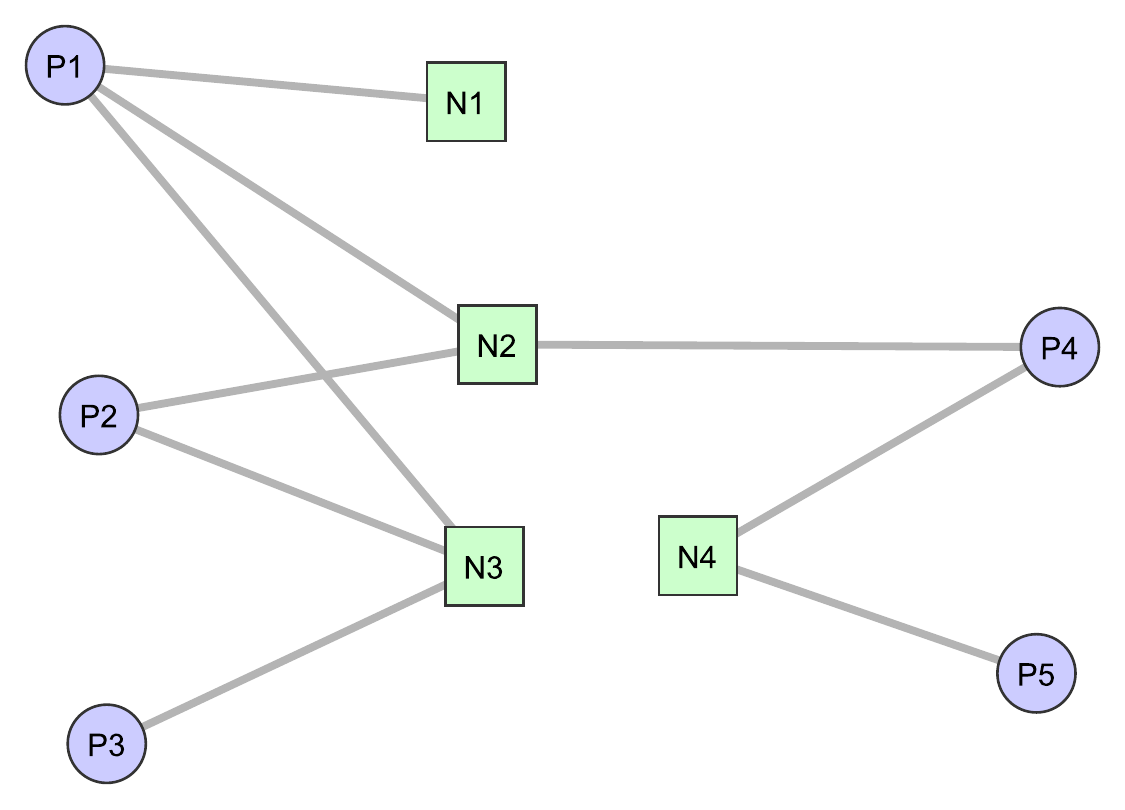}
\end{SCfigure}

To see why we think paths as well as direct link may have meaning in the context of our enhancement network, we think it is helpful to look at a trivial example. Consider the perfume-note network $\Gcal_\mathrm{ex}$ given in \figref{fpnex} where
\begin{itemize}
  \item P1 (perfume 1) contains notes N1, N2 and N3,
  \item P2 contains notes N2, and N3,
  \item P3 contains notes N3,
  \item P4 contains notes N2 and N4,
  \item P5 contains note N4,
  \item P1 has more votes than P2, which has more votes than P3,
  \item P4 has more votes that P5.
\end{itemize}

\begin{SCfigure}[50][htb!]
  \centering
  \caption{The enhancement network $\Hcal_\mathrm{ex}$ derived from the perfume-note network $\Gcal_\mathrm{ex}$  shown in Figure 1 where the perfume votes are as indicated in the caption of \figref{fpnex}.}
  \label{fenetex}
  \includegraphics[width=0.3\linewidth]{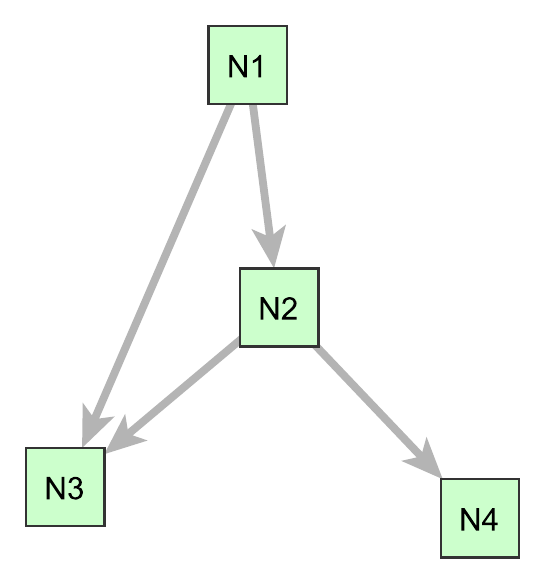}
\end{SCfigure}

Then the enhancement network $\Hcal_\mathrm{ex}$ derived from $\Gcal_\mathrm{ex}$  contains the following directed links
\begin{itemize}
  \item from N1 to N2 (from P1 overlap with P2),
  \item from N1 to N3 (from P1 overlaps with P2 and P3),
  \item from N2 to N3 (from P2 overlap with P3),
  \item from N2 to N4 (from P4 overlap with P4).
\end{itemize}
This is illustrated in \figref{fenetex}.

Hence there is a two-step directed path $N1 \rightarrow N2 \rightarrow N3$.  Here this two-step path N1 to N2 to N3 does imply some sort of relationship between the ends, between N1 and N3. After all there is also a direct link from N1 to N3 derived from the overlap of P1 and P3. This direct link will mean N3 does contribute strongly to the centrality of N1 though the indirect link will be a smaller contribution to the centrality score of N1.

Now there is also a second two-step directed path, from N1 to N2 to N4. However, in this case there is no direct edge from N1 to N4. Without a direct link, N4 will not contribute so strongly to the centrality of N1 but it be a small contribution.

The interesting point comes when we consider the importance of note N4 in any perfumes containing note N1.  As with any recipe, the precise combination matters, the whole ``is more than the sum of its parts''. There is no direct evidence here that N1 will enhance N4 which is reflected in the lack of a direct link. This is, however, the the point of a network representation.  The indirect relationship as captured by a path is some sort of indication, a suggestion or recommendation, that is a direct link between the ends of a path is likely to be a good idea.  Consider a musical example, which is another nice example of a recipe. If N3 and N4 are musical notes an octave apart, or if N3 and N4 are the same musical note played by different instruments in a score, then it is a good suggestion to try notes N1 and N4 in a chord since N1 and N2 work well as musical chords.

Of course, we have no proof that indirect links are good recommendations. In our musical example, a chord N1 and N4 might be terrible because the instrument playing note N4 rather than N2 just clashes with the instrument playing note N1. The lack of a direct connection may also be a recommendation in itself, no one has made that connection for a good reason. What we are aiming to do in our analysis, as always in networks, is to look for many such indirect suggestions.  Then even if some paths represent poor suggested relationships, we hope that with many such suggestions, through the network analysis of the many possible paths in the network, the weight of good indirect relationships will reinforce each other in the way that many more poor combinations are unlikely to do.

So we assume that paths in our enhancement network can be a useful tool, just as paths rather than direct links are important in any network context. Clearly, if recipes are more than the sum of their parts, one might feel that in such contexts (music scores, perfume and food recipes etc) higher order effects (non-backtracking matrices, hypergraphs, clique overlaps etc) might provide more effective insights than simple bilateral relationships recorded in ordinary graphs.  But then, that is exactly why we developed our enhancement network; a traditional network encoding higher order effects (a type of clique overlap) in the original bipartite perfume-note graph.

Incidentally, this discussion of the enhancement network and its paths also reinforces an argument why making a projection onto a network of notes alone is worthwhile.  In making this one-mode projection we undoubtedly lose some information.  However, the way we do the projection to create the enhancement network highlights key higher-order network effects in the original perfume-note network, higher order effects which can be drown out in standard measurements made on the original perfume-note network.

Once we have decided that paths as well as direct links in the enhancement network contain useful information then it is natural to use traditional centrality measures on the enhancement network.




\subsubsection{Enhancement Network Centrality Measures}\label{ssenetcent}

We used centrality measures to analyse the enhancement network. Centrality is a measure of importance of nodes, given their connections and position within a network. \citet{N10} explains in detail the majority of main centrality measures. For our purposes, we used \textit{weighted out-degree} centrality, \textit{out-closeness} and \textit{reversed PageRank}, defined as follows.

\textit{Weighted out-degree} $s_n$ (strength) of a note $n$ in the enhancement network is
\begin{equation}
s_n = \sum_m H_{mn} \, .
\end{equation}
It is equal to the size of the multiset of notes directly enhanced by note $n$.

We use \textit{out-closeness}, $c_n$ for note $n$, in the form proposed by \cite{WF94}. Here we count the outward going paths as this captures chains of enhancement.
\begin{equation}
 c_n =
 \frac{|\Ecal_n|}{(N-1)} \frac{|\Ecal_n|}{f_n} \, ,
 \qquad
 f_n = \sum_{m \in \Ecal_n} d_{mn}
 \, .
\end{equation}
We define $\Ecal_n$ to be the nodes which are reachable from node $n$ (i.e.\ notes directly and indirectly enhanced by note $n$). So if $m \in \Ecal_n$ there is a directed path from node $n$ to node $m$.
For these directed paths, we define the length of each  path to be the sum of the weights along that path.  Then $d_{mn}$ is the length of the shortest path from $n$ to $m$. The factor $1/(N-1)$ is an irrelevant overall constant but for completeness we note that here we have used $N$ to be the number of nodes (notes) in the LWCC (largest weakly connected component) of the enhancement network.
A large closeness centrality score is assigned to nodes that have short paths to many nodes in the network. Thus, out-closeness centrality evaluates the universal note’s potential to enhance: if a note enhances many notes, multiple times, it is assigned a large score.

We also studied \textit{reversed PageRank}, or PageRank on the enhancement network with reversed edges.  For node $n$ it is defined to be $\PR_n$ where
\begin{equation}
 \PR_n =
 \frac{1}{N}(1-\alpha) +
 \alpha \sum_{m} \frac{1}{k_m^{\textrm{in}}} H_{mn} \, \PR_m \, ,
 \qquad
 k_m^{\textrm{in}} = \sum_{m}  H_{mn} \, .
\end{equation}
PageRank of note $n$ counts the number notes it enhances but weights each enhanced neighbour by its importance, its PageRank. So the larger the PageRank of a note, the more important it is in terms of enhancing other notes. PageRank can be understood as a  random walk where the walkers leave nodes $m$ along incoming edges and the probability of a random walker following an incoming edge to move from $m$ to a neighbouring node $n$, which is enhancing $m$, is proportional to the weight of the edge multiplied by $\alpha$. There is also a second process, which occurs with probability $(1-\alpha)$, where the random walker starts again from a node chosen uniformally at random from the set of nodes in the graph (a `hyperjunp'). The probability of finding a random walker at a node in the long-time limit is the PageRank score for that node.

Several values for $\alpha$ were tried but we found little difference for $\alpha$ between $0.7$ and $0.95$. Therefore, we used the value $\alpha=0.85$ which corresponds to the random walkers making on average $5.67$ steps before a hyperjump compared to the the average shortest path length of $2.9$.

\end{document}